\documentclass[11pt,a4paper]{article}

\pdfoutput=1

\usepackage{amsthm,amsmath}
\usepackage[charter]{mathdesign}
\usepackage{isomath}

\usepackage[top=3.2cm,bottom=3.4cm,left=3.2cm,right=3.2cm]{geometry}
\usepackage[round]{natbib}%
\usepackage{xcolor}
\usepackage{hyperref}
\usepackage[utf8]{inputenc}
\usepackage[OT1]{fontenc}
\usepackage[english]{babel}
\usepackage{graphicx}
\usepackage[perpage,symbol]{footmisc}
\usepackage[hang,labelfont=bf,size=small]{caption}
\usepackage[protrusion=true,expansion=true]{microtype}

\hypersetup{
  colorlinks,
  urlcolor=black,
  citecolor=black,
  linkcolor=black,
}

\title{FreeSASA: An open source C library for solvent accessible surface
  area calculations
}
\author{Simon Mitternacht}

\begin{document}
\thispagestyle{empty}
\begin{flushright}
January 25, 2016
\end{flushright}\vspace*{1cm}
\begin{center}
\noindent
{\huge\bfseries FreeSASA: An open source C library for\\\medskip
   solvent accessible surface area calculations}\\\bigskip\bigskip
{\large Simon Mitternacht}\\\bigskip
\emph{University Library, University of Bergen}\\
\emph{Postboks 7808, 5008 Bergen, Norway}\\\medskip
\texttt{simon.mitternacht@uib.no}\\\bigskip
\end{center}
\begin{abstract}
  \noindent Calculating solvent accessible surface areas (SASA) is a
  run-of-the-mill calculation in structural biology. Although there
  are many programs available for this calculation, there are no
  free-standing, open-source tools designed for easy tool-chain
  integration. FreeSASA is an open source C library for SASA
  calculations that provides both command-line and Python interfaces
  in addition to its C API. The library implements both Lee and
  Richards' and Shrake and Rupley's approximations, and is highly
  configurable to allow the user to control molecular parameters,
  accuracy and output granularity. It only depends on standard C
  libraries and should therefore be easy to compile and install on any
  platform. The source code is freely available from
  \url{http://freesasa.github.io/}.  The library is well-documented,
  stable and efficient. The command-line interface can easily replace
  closed source legacy programs, with comparable or better accuracy
  and speed, and with some added functionality.\bigskip\\
  \textbf{Abbreviations}: API: Application programming
  interface. SASA: Solvent accessible surface area. L\&R: Lee and
  Richards' approximation. S\&R: Shrake and Rupley's approximation.

  \begin{description}
  \item[Home page:] \url{http://freesasa.github.io/}
  \item[License:] GNU GPL 3.0.
  \item[Programming languages:] C, Python.
  \item[Dependencies:] Standard C and GNU libraries,
    C99-compliant compiler. Cython to build Python bindings
    (optional).
  \end{description}
\end{abstract}
\newpage

\section*{Background}
Exposing apolar molecules to water is highly unfavorable, and
minimizing the hydrophobic free energy is an important driving force
in the folding of macromolecules \citep{finkelstein2002protein}. The
solvent accessible surface area (SASA) of a molecule gives a measure
of the contact area between molecule and solvent. Although the exact
quantitative relation between surface area and free energy is elusive,
the SASA can be used to compare different molecules or different
conformations of the same molecule, or for example measure the surface
that is buried due to oligomerization.

To define the SASA, let a spherical probe represent a solvent
molecule. Roll the probe over the surface of a larger molecule. The
surface area traced by the center of the probe is the SASA of the
larger molecule \citep{LnR}. Two classical approximations are commonly
used to calculate SASA: one by Lee and Richards (L\&R) where the
surface is approximated by the outline of a set of slices
\citeyearpar{LnR}, and one by Shrake and Rupley (S\&R) where the
surface of each sphere is approximated by a set of test points
\citeyearpar{SnR}. The SASA can be calculated to arbitrary precision
by refining the resolution of both. The surface area can also be
calculated analytically \citep{fraczkiewicz1998exact}, which is useful
when the gradient is needed, or by various other approximations,
tailored for different purposes \citep{sanner1996reduced,
  weiser1999approximate, POPS, xu2009generating,
  drechsel2014triforce}.

There are many tools available to calculate SASA. The most popular
program for command line use is probably NACCESS \citep{NACCESS}
(freely available for academic use), which is an efficient Fortran
implementation of the L\&R approximation. Another well-known command
line tool is DSSP, which primarily calculates the secondary structure
and hydrogen bonds of a protein structure, but provides the SASA as
well \citep{DSSP} (using S\&R, open source). There are also some web
services available, for example Getarea, which calculates the surface
analytically \citep{fraczkiewicz1998exact}, and Triforce which uses a
semianalytical tessellation \citep{drechsel2014triforce} (also
available for command line use). In addition, most molecular dynamics
simulation packages include tools to analyze SASA from trajectories.

FreeSASA is intended to fill the same niche as NACCESS, and a number
of other similar programs: a simple and fast command-line tool that
``does one thing and does it well'' and can be easily integrated into
tool chains. The advantage of FreeSASA is that it is open source (GNU
General Public License 3), and provides both C and Python APIs in
addition to its command line interface. It has sensible default
parameters and no obligatory configuration for casual users (the only
required input is a structure), but also allows full control over all
calculation parameters. Dependencies have been kept to a minimum:
compilation only requires standard C and GNU libraries. The library is
thread-safe, and some effort has gone into dealing gracefully with
various errors. The code ships with thorough documentation, which is
also available online at
\url{http://freesasa.github.io/doxygen/}. Although functionality and
availability have been the primary motivating factors for writing this
library, performance tests show that FreeSASA is as fast as or faster
than legacy programs when run on a single CPU core. In addition, a
large portion of the calculation has been parallelized, which gives
significant additional speed when run on multicore processors.

\section*{Implementation}

\subsection*{Calculations}

Both S\&R and L\&R are pretty straightforward to implement, and both
require first determining which atoms are in contact, and then
calculating the overlap between each atom and its neighbors. Finding
contacts is done using cell lists, which means the contact calculation
is an $O(N)$ operation. Both algorithms then treat each atom
independently, making also the second part of the calculation
$O(N)$. In addition, this second part is trivially parallelizable.

For L\&R, instead of slicing the whole protein in one go, each atom is
sliced individually. The L\&R calculation is thus parameterized by the
number of slices per atom, i.e.\ small atoms have
thinner slices than large atoms.

The Fibonacci spiral gives a good approximation to an even
distribution of points on the sphere \citep{FibonacciGrid}, allowing
efficient generation of an arbitrary number of S\&R test points. The
cell lists provide the first of the two lattices in the double cubic
lattice optimization for this algorithm \citep{DCLM}, the second
lattice (for the test points) is not implemented in FreeSASA, for now.

The correctness of the implementations was tested by first inspecting
the surfaces visually. In the two atom case, results were verified
against analytical calculations. Another verification came from
comparing the results of high precision SASA calculations using the
two independent algorithms. In addition, using the L\&R algorithm
gives identical results to NACCESS when the same resolution and atomic
radii are used.

\subsection*{Radius assignment}
An important step of the calculation is assigning a radius to each
atom. The default in FreeSASA is to use the \emph{ProtOr} radii by
\citet{tsai1999packing}. The library recognizes the 20
standard amino acids (plus Sec and Pyl), and the standard nucleotides
(plus a few nonstandard ones). Tsai et al.\ do not mention Phosphorus
and Selenium; these atoms are assigned a radius of 1.8 and 1.9
\AA\ respectively. 

By default, hydrogen atoms and HETATM records are ignored. If
included, the library recognizes three common HETATM entries: the
acetyl and $\mathrm{NH_2}$ capping groups, and water, and assigns
ProtOr radii to these. Otherwise the van der Waals radius of the
element is used, taken from the paper by
\citet{mantina2009consistent}. For elements outside of the 44 main
group elements treated by Mantina et al., or if completely different
radii are desired, users can provide their own configuration.

Users can specify their own atomic radii either through the API or by
providing a configuration file. The library ships with a few sample
configuration files, including one that provides a subset of the
NACCESS parameterization, and one with the default ProtOr
parameters. In addition, scripts are provided to automatically
generate ProtOr configurations from PDB CONECT entries, such as those
in the Chemical Component Dictionary
\citep{westbrook2014chemical}. These can then be appended to the
default configuration.

\subsection*{Command-line interface}

Building FreeSASA creates the binary \verb|freesasa|. The simplest
program call, with default parameters, is
\begin{verbatim}
   $ freesasa 3wbm.pdb
\end{verbatim}
using the structure with PDB code 3wbm as example (a protein-RNA
complex). The above produces the following output
\begin{verbatim}
   ## freesasa 1.0 ##

   PARAMETERS
   algorithm    : Lee & Richards
   probe-radius : 1.400
   threads      : 2
   slices       : 20

   INPUT
   source  : 3wbm.pdb
   chains  : ABCDXY
   atoms   : 3714

   RESULTS (A^2)
   Total   :   25190.77
   Apolar  :   11552.38
   Polar   :   13638.39
   CHAIN A :    3785.49
   CHAIN B :    4342.33
   CHAIN C :    3961.12
   CHAIN D :    4904.30
   CHAIN X :    4156.46
   CHAIN Y :    4041.08
\end{verbatim}
The numbers in the results section are the SASA values (in
$\text{\AA}^2$) for the respective groups of atoms. 

As an illustration of a few of the other configuration options, the
command
\begin{verbatim}
   $ freesasa -n 100 --print-as-B-values --no-log < 3wbm.pdb > 3wbm.sasa
\end{verbatim}
calculates the SASA of a PDB-file passed via \verb|stdin|, using 100
slices per atom. The flag \verb|--no-log| suppresses the regular
output. The output will instead, because of the flag
\verb|--print-as-B-values|, be the provided PDB-file with the SASA of
each atom replacing the temperature factors, and the atomic radii
stored in the occupancy factor field. The program can thus be used as
a PDB-file filter. 

The command-line interface also provides facilities to analyze
individual chains or groups of chains in a structure separately.  By
calling with the option \verb|--chain-groups|,
\begin{verbatim}
   $ freesasa --chain-groups=ABCD+XY 3wbm.pdb
\end{verbatim}
we get an output similar to the first example, but with two additional
calculations, one where only protein chains A, B, C and D have been
included, and one with only the RNA chains X and Y. 

The option \verb|--select| can be used to select a set of atoms using
a subset of the syntax used in the program Pymol
\citep{delano2002pymol}. For example, the command
\begin{verbatim}
   $ freesasa --select="RNA, resn A+U+G+C"
\end{verbatim}
will produce the following output (after the regular output shown
above)
\begin{verbatim}
   SELECTIONS
   RNA :    8197.53
\end{verbatim}
where \emph{RNA} is simply the user-defined name of the selection, and
the number the contribution to the total SASA from the bases A, U, G
and C (which we in this particular case could have got by simply
adding the areas for the chains X and Y above).

The command
\begin{verbatim}
   $ freesasa -h
\end{verbatim}
prints a help message listing all available options, including other
ways to redirect output and how to change different calculation
parameters (the most detailed information can be found online at
\url{http://freesasa.github.io/doxygen/CLI.html}).

\subsection*{C API}

The C code below illustrates how to perform a SASA-calculation on the
same PDB-file as above, using the C API, with default parameters. The
functions and types used are all defined in the header
\verb|freesasa.h|.
\begin{verbatim}
    FILE *fp = fopen("3wbm.pdb","r");
    freesasa_structure *structure
        = freesasa_structure_from_pdb(fp, NULL, 0);
    freesasa_result *result
        = freesasa_calc_structure(structure, NULL);
    printf("Total area : %f A2\n", result->total);
\end{verbatim}
The two points where null pointers are passed as arguments are places
where atom classifiers and calculation parameters could have been
provided. A more elaborate example, that includes error checking and
freeing of resources is included in the appendix.

The API also allows the user to calculate the SASA of a set of
coordinates with associated radii. The code below puts two atoms at
positions $\vec{x}_1 = (1,1,1)$ and $\vec{x}_2 = (2,2,2)$ with radii
$r_1 = 2$ and $r_2 = 3$, respectively, and
outputs the SASA of the individual atoms.
\begin{verbatim}
   double coord[] = {1.0, 1.0, 1.0, 2.0, 2.0, 2.0};
   double radius[] = {2.0, 3.0};
   freesasa_result *result = 
       freesasa_calc_coord(coord, radius, 2, NULL);
   printf("A1 = %f, A2 = %f\n", result->sasa[0], 
          result->sasa[1]);
\end{verbatim}

\subsection*{Python API}
The library includes Python bindings that export most of the C API to
Python. The Python code below gives the same output as the first C
example, plus the area of polar and apolar atoms. Again, error
handling is excluded for brevity.

\begin{verbatim}
   import freesasa

   structure = freesasa.Structure("3wbm.pdb")
   result = freesasa.calc(structure)
   classArea = freesasa.classifyResults(result,structure)

   print "Total : %.2f A2" % result.totalArea()
   for key in classArea:
       print key, ": %.2f A2" % classArea[key]
\end{verbatim}

\section*{Performance}

The computational efficiency of the two algorithms was compared by
running the FreeSASA command-line program with different parameters on
a representative set of PDB structures. This set was generated from
the most restrictive list of structures in the PISCES database
\citep{PISCES}. 88 PDB files were selected randomly from a set of size
intervals in this list, to get an approximately even distribution in
size\footnote{Resulting in the following set of proteins: 1d8w, 1fo8,
  1g5a, 1gqi, 1h16, 1hbn, 1hbn, 1jcd, 1jz8, 1kqf, 1m15, 1mj5, 1n62,
  1o6g, 1oew, 1q6z, 1rk6, 1rwh, 1su8, 1t1u, 2bz1, 2cvi, 2d5w, 2drt,
  2dvm, 2e4t, 2e7z, 2eab, 2gb4, 2gj4, 2gpi, 2heu, 2hhc, 2i53, 2odk,
  2olr, 2qud, 2yfo, 2zux, 3a5f, 3b0x, 3bon, 3bvx, 3cuz, 3fo3, 3fss,
  3g02, 3hkw, 3i94, 3ie7, 3kyz, 3lqb, 3m0z, 3og2, 3pzw, 3sqz, 3t47,
  3tew, 3tg7, 3u9w, 3uc7, 3w7t, 3w7y, 3wa2, 3weo, 3whr, 3znv, 3zsj,
  4a9v, 4bb9, 4bps, 4btv, 4c1a, 4cj0, 4g6t, 4g7x, 4gmu, 4h7u, 4kv7,
  4la2, 4lix, 4n13, 4oh7, 4oxx, 4p0t, 4pj2, 4qas, 4xfj, 7odc.}. PISCES
specifies a specific chain in each structure, but in the following all
chains were used, which resulted in the largest structure having over
30\,000 atoms (1jz8). To average out some variation in the running
time in these rather short calculations (in some cases $<
10\,\mathrm{ms}$), the fastest calculations were run two to five
times. As we will see below, error bars are relatively small along
that axis.

To measure the accuracy of the two algorithms, a reference SASA value,
$A_\text{ref}$, was calculated using L\&R with 1000 slices per atom
for each structure. The error of a given SASA-value, $A$, is then
$\varepsilon = \lvert A - A_\text{ref} \rvert / N$, where $N$ is the
number of atoms in the structure.

\begin{figure}
  \begin{center}
  \includegraphics{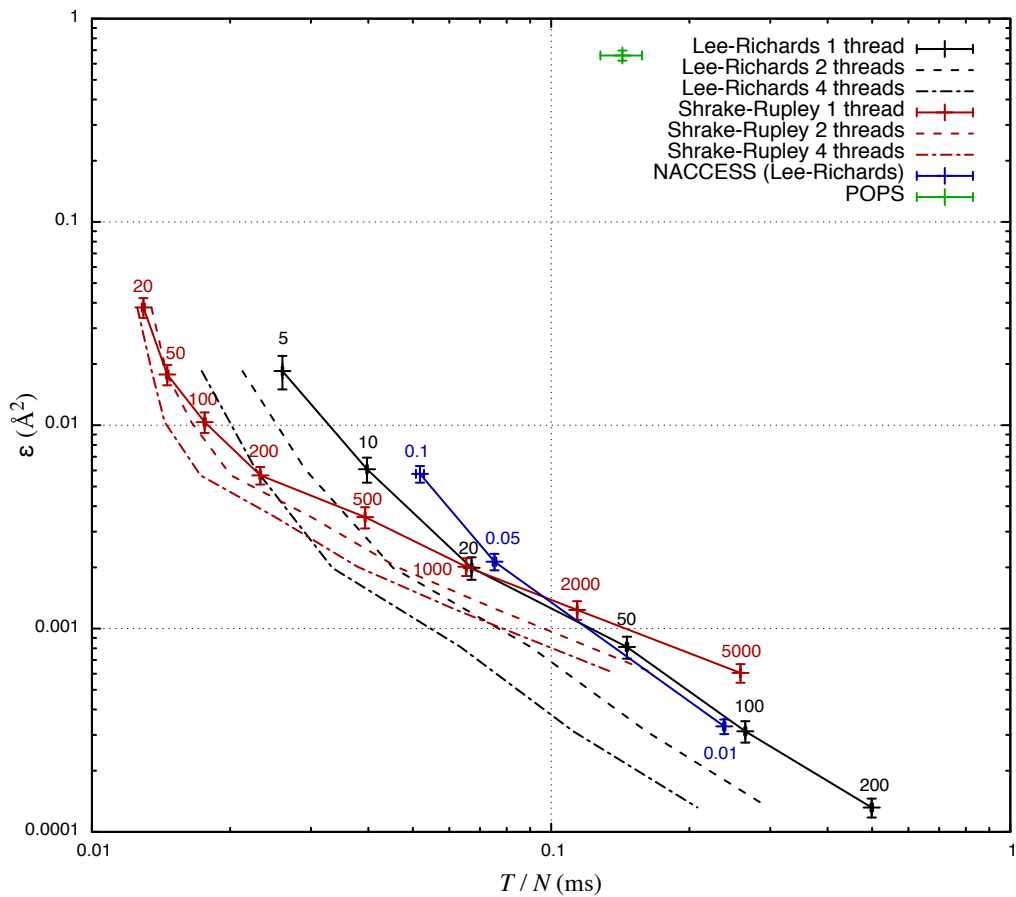}
  \caption{Precision and calculation time. The mean of the error
    $\varepsilon$ in SASA vs calculation time per atom $T/N$, for the
    two algorithms in FreeSASA plus the programs NACCESS and
    POPS. Labels indicate the resolution used for each set of
    calculations, and error bars the standard error along both
    axes. The solid lines are only there to guide the eye, and the
    dashed lines indicate the analogous lines when using 2 and 4
    threads in FreeSASA. An L\&R run with 1000 slices was used as
    $A_\text{ref}$ when calculating $\varepsilon$ for both
    approximations. NACCESS uses L\&R and was run with three values of
    the z-parameter (0.1, 0.05 and 0.01, corresponding to 10, 20 and
    100 slices per atom), a run with z-parameter 0.005 was used as
    $A_\text{ref}$ (using even lower z-values gave inconsistent
    results). The NACCESS reference calculation was also used as
    reference for POPS. All programs were compiled using GCC 4.9.3
    with the optimization flag ``-Ofast'' and the tests were run on an
    Intel Core i5-2415M CPU at 2.30\,GHz.\label{fig:precision}}
  \end{center}
\end{figure}

Figure~\ref{fig:precision} shows the results of these calculations for
the 88 proteins described above. At low resolution S\&R is
considerably faster than L\&R, and at high resolution L\&R is faster,
with a crossover at 1000 test points or 20 slices per atom (20 slices
is the default setting in FreeSASA).

Comparisons were done with NACCESS \citep{NACCESS}, DSSP \citep{DSSP},
NSOL 1.7 \citep{masuya2003nsol}, POPS \citep{POPS} and Triforce
\citep{drechsel2014triforce}. The list could potentially have been a
lot longer; some programs were left out on the basis of being
closed source, poorly documented or no longer available. NACCESS was
included in spite of its limiting license due to its popularity. The
SASA facilities in molecular dynamics packages were not considered
since these cater to a different use case.

NACCESS allows the user to choose arbitrary precision and can
therefore be used as a reference for itself, and POPS was optimized
with NACCESS as reference. NACCESS uses L\&R and performs very
similarly to FreeSASA using L\&R. The POPS algorithm is intended as a
fast coarse-grained approximation; its authors state an average error
of $2.6\ \text{\AA}^2$ per atom \citep{POPS}. In
figure~\ref{fig:precision} the mean $\varepsilon$ is lower than that,
which is expected, since this error is measured over the total SASA,
not atom by atom. A fit showed that POPS runs in $O(N^2)$ time (data
not shown), which to some extent explains the relatively long mean
calculation time per atom.

The other programs listed above were left out of figure
\ref{fig:precision} because they can not be compared under the same
premises. DSSP calculates many different things in addition to its 200
test-point S\&R-calculation, and the total running time is therefore
naturally longer than the corresponding calculation in FreeSASA,
although the accuracy should be comparable for the same number of test
points. The program NSOL uses S\&R, but does five different SASA
calculations on the same input using different parameters. The NSOL
code was modified to only do one of the five calculations, and is then
only slightly slower than FreeSASA using the same number of test
points. Lastly, Triforce is not suitable for comparison in this
particular use case because it has a high initialization cost, which
makes it slow for calculating the SASA of an isolated structure.

\begin{figure}
  \begin{center}
  \includegraphics{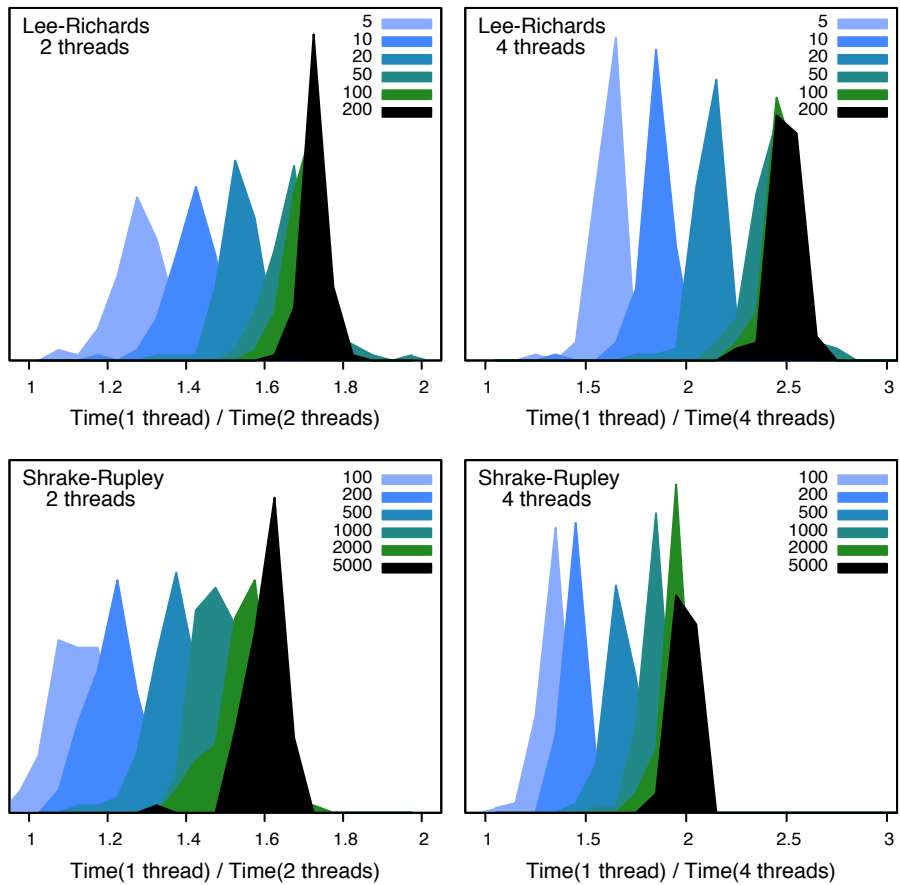} 
  \caption{Parallelization. The histograms shows the
    distribution of the calculation time using two or four threads
    divided by the time using one thread. Thus if this fraction is two
    or four, respectively, we have ``perfect'' parallelization.  The
    legends indicate the resolution of the calculation: for L\&R,
    slices per atom, and for S\&R, number of test points.
    \label{fig:threads}}
  \end{center}
\end{figure}

In single-threaded mode, FreeSASA using L\&R is almost
indistinguishable from NACCESS in figure~\ref{fig:precision}, but it
is significantly faster when 2 or 4 threads are used. The effect of
spreading the calculation over several threads is shown in more detail
in figure~\ref{fig:threads}. Since the generation of cell lists is not
parallelized, using more than one thread only gives a significant
performance benefit in the high resolution limit. Based on these
results, the default has been set to two threads. Depending on the
nature of the calculations, this speedup can make a noticeable
difference.

\section*{Summary}
FreeSASA is an efficient library for calculating the SASA of protein,
RNA and DNA structures. Its main advantages over other commonly used
tools is that it is open source, easily configurable and can be used
both as a command line tool, a C library and a Python module. The
tests above demonstrate that it runs as fast as, or faster than, some
popular tools at a given resolution, and can be boosted further by
parallelizing the calculation.

\begin{small}

\subsection*{Acknowledgments}
Thanks go to Edvin Fuglebakk for comments on the code and
documentation. Thanks to Sandhya Tiwari and Anders Irb\"ack for
comments on the manuscript. Thanks to Jo\~ao Rodrigues for suggestions
for improved functionality.

\bibliographystyle{plainnat}

\end{small}

\section*{Appendix}
Below is an example C program, illustrating how to use the C API for a
basic SASA calculation on a structure read from \verb|stdin|,
including error checking and freeing of resources, using default
parameters.
\begin{verbatim}
    #include <stdlib.h>
    #include <stdio.h>
    #include "freesasa.h"

    int main(int argc, char **argv) {
        freesasa_structure *structure = NULL;
        freesasa_result *result = NULL;
        freesasa_strvp *class_area = NULL;

        /* Read structure from stdin */
        structure = freesasa_structure_from_pdb(stdin,NULL,0);

        /* Calculate SASA using structure */
        if (structure) {
            result = freesasa_calc_structure(structure,NULL);
        }

        /* Calculate area of classes (Polar/Apolar/..) */
        if (result) {
            class_area = freesasa_result_classify(result,structure,NULL);
        }

        /* Print results */
        if (class_area) {
            printf("Total area : %f A2\n",result->total);
            for (int i = 0; i < class_area->n; ++i)
                printf("%s : %f A2\n",class_area->string[i],
                       class_area->value[i]);
        } else {
            /* If there was an error at any step, we will end up here */
            printf("Error calculating SASA\n");
        }

        /* Free allocated resources */
        freesasa_strvp_free(class_area);
        freesasa_result_free(result);
        freesasa_structure_free(structure);

        return EXIT_SUCCESS;
    }
\end{verbatim}
\end{document}